\documentstyle[12pt]{article}
\begin{document}
\thispagestyle{empty}

\hfill SNUTP-93-22

\vskip 1.5cm
\begin{center}
{\Large\bf Dark Matters in Axino Gravitino Cosmology}
\vskip 2cm
Eung Jin Chun$^a$, Hang Bae Kim$^{a,b}$ and Jihn E. Kim$^{a,b}$

\sl Center for Theoretical Physics$^a$ and
Department of Physics$^b$\\
Seoul National University\\
Seoul 151-742, Korea
\rm
\end{center}
\vskip 3.5cm
{\center Abstract\\}

It is suggested that the axino mass in the 1 MeV region and
gravitino mass in the eV region
can provide an axino lifetime of
order of the time of photon decoupling.  In this case,
some undecayed axinos act like cold dark matters and some axino
decay products (gravitinos and hot axions) act like hot
dark matters at the time of galaxy formation.

\newpage

The cosmological constant problem  and the dark matter problem
seem to be the two most pressing issues in current cosmology.
The ultimate solution of the cosmological constant
problem may need a new theory of particle physics.  On the other
hand, the dark matter problem is simply {\it $\lq\lq$What is the
dark matter?"}  A small cosmological constant
in particle theory may act like a dark matter in some sense,
but we will not
dwell on this possibility in this paper in view of no consistent
solution of the cosmological constant problem.

Thus the widely discussed candidates for the dark matter are
centered on elementary particles, which are grouped to {\it hot,
warm, and cold dark matters} depending on their kinetic energies at
the time of matter--radiation equality \cite{kolb}.
For neutrinos, 10 eV neutrino is
hot dark matter, keV neutrino is warm dark matter, and GeV
neutrino is cold dark matter.   For neutrinos, it is relatively
easy to test the cosmological idea in particle physics experiments.
The unknown parameters of neutrinos are simply their masses
\cite{data}:
$m_{\nu_e}<7.3\ {\rm eV},\ m_{\nu_\mu}<0.27\ {\rm MeV},\ m_{\nu
_\tau}<35\ {\rm MeV}$.  Other dark matter candidates such as axion,
photino or neutralino (photino and zino mixed with neutral Higgsinos)
are cold dark matters.  These
latter candidates are not discovered yet, but the theoretical
motivation for these particles and their expected role in
cosmology are so immense that they have attracted
a great deal of attention
recently.  For axions, even the low temperature cavity detectors
have been built under the assumption that axions are the dark
matter of the universe \cite{sikivie}.
The photino candidate belongs to the
so-called WIMP (weakly interacting massive particle) category
\cite{wimp}.

One progress in particle physics is gradual acceptance of supersymmetry
\cite{kane}
as the solution of the gauge hierarchy problem (or the scalar mass
problem).  In most supersymmetric theories, the $R$-symmetry dictates
the existence of a
stable particle which is commonly called LSP (the lightest supersymmetric
particle).  In the literature, the LSP cosmology (mostly photino or
neutralino) has been extensively discussed \cite{ellis}.
We will not consider the
possibility of $R$ violating interactions in which case there is no
LSP \cite{hall}.

In this paper, we concentrate on the axino--gravitino cosmology.
For this, we need
a Peccei--Quinn symmetry with supersymmetry.  The Peccei--Quinn symmetry
is needed for a solution of the strong CP problem \cite{cp},
and the supersymmetry is introduced for a solution of the gauge hierarchy
problem \cite{nilles}.  In this
theory, the cosmological effects of the axino (the superpartner of axion)
and the gravitino (the superpartner of graviton) are important because
their interactions are feeble due to the suppression factor of the
axion decay constant or the Planck mass.  Axino ($\equiv\tilde a$)
cosmology \cite{axino} and gravitino ($\equiv\tilde G$)
cosmology \cite{gravitino} have been discussed previously.
Currently favored gravitino mass is identified with the
electroweak scale
of 100--1000 GeV.  These heavy gravitinos decay and cannot be a
candidate of the dark matter, but they might have affected the deuterium
abundance \cite{ekn}.  Recently, however, the old idea of supercolor
at multi TeV region has been reinvestigated without
phenomenological difficulties
\cite{dine}.  In this case the gravitino mass falls in the $0.4\times
10^{-3}$ eV -- 0.4 keV region for $\Lambda_{\rm supercolor}$ =
1--$10^3$ TeV.
Therefore, it is worthwhile to consider the cosmological
consequence of a light gravitino. We assume that {\it
this light gravitino is the LSP.}  In Ref. \cite{dine}, the axion is
made heavy by introducing an $R$--color, not introducing the $\mu H_1
H_2$ term.  Here, we introduce the
Peccei--Quinn symmetry at $10^{12}$ GeV and the
$\mu$--term explicitly, and study the
axino--gravitino cosmology with a light gravitino.

Gravitinos decouple just after the Planck time and a keV gravitino
can be a candidate for the dark matter in the standard Big Bang
cosmology.  However, the needed
inflation after the Planckian epoch might have washed out the
gravitinos almost completely.  Nevertheless, the keV gravitinos can
be produced by the decay of SLSPs
(the second lightest supersymmetric particles).  If the SLSP is
photino ($\equiv\tilde\gamma$),
the photino decay $\tilde\gamma\rightarrow \tilde G +\gamma$
is the dominant decay mode.
However, the number density of photino at the decoupling is
very small compared to the photon number density,
\begin{equation}
{n_{\tilde\gamma}\over n_\gamma}\Big|_{T_{\tilde\gamma D}}\simeq
10^{-9}
\end{equation}
where $T_{\tilde\gamma D}= m_{\tilde\gamma}/20$ is used.
In this case, even a keV gravitino cannot serve as a dark matter.

To produce enough gravitinos from the SLSP, the SLSP must decouple at
temperature above the scale of its mass so that the Boltzmann
suppression is not present as in the photino case.  Then the best
candidate for the SLSP seems to be the axino $\tilde a$.  In addition,
the decay products of axinos are axions and gravitinos which cannot
affect the precious deuterium abundance.
Futhermore, the axino has the
well-known structure for its interaction with other particles,
suppressed by the axion decay constant $F_a$.  The allowed axion
decay constant $F_a\simeq 10^{10}\sim 10^{12}$ GeV is huge compared
to the weak scale, and hence the lifetime of the axino can fall in
the cosmologically interesting period around the time of galaxy
formation.

To study this axino--gravitino cosmology in the above
scenario, we assume the LSP is the gravitino $\tilde G$ ($m_{
3/2}\sim$ eV), the SLSP is the axino $\tilde a$ ($m_{\tilde a}
\sim$ 1 MeV), and the LOSP (lightest ordinary supersymmetric
particle) is the photino.  The interesting mass ranges of these
particles turn out to be $m_{3/2}\sim $ eV, $m_{\tilde a}=1$
MeV, and $m_{\tilde \gamma}>$ 20 GeV.

The lifetimes of the photino and the axino are
\begin{equation}
\tau_{\tilde\gamma}\ =\ {4\pi F_a^2
\over 3C^2_{a\gamma\gamma} m^3_{\tilde
\gamma}}\left(1-{m_{\tilde a}^2\over m^2_{\tilde\gamma}}\right)^{-3}
\end{equation}
\begin{equation}
\tau_{\tilde a} ={96\pi M^2 m_{3/2}^2\over m_{\tilde a}^5}\ =\ 1.2\times
10^{12}\ {\rm sec}\left({{\rm MeV}\over m_{\tilde a}}\right)^5
\left(m_{3/2}\over {\rm eV}\right)^2
\end{equation}
where $C_{a\gamma\gamma}$ is the
axion--photon coupling defined in Ref. \cite{cp}.
For the axino decay we use the interaction \cite{sugra},
\begin{equation}
{1\over M}\bar\psi_\mu \gamma^\nu\partial_\nu z^{*}\gamma^\mu
\tilde a_L+(\rm h.c.)
\end{equation}
where $z=(s+ia)/\sqrt{2}$ in terms of {\it saxino} $s$ and axion $a$.
There exists another interaction term
proportional to $m_{3/2}$ which is negligible compared to the
above interaction.
The axino
lifetime can fall in the region of photon decoupling.

For supergravity interactions, the squared $T$--matrix contains
gravitino mass in the denominator due to the $1/m_{3/2}^2$ term
in the tensor structure  of the gravitino polarization
\begin{equation}
P_{\rho\sigma}=(\gamma_\mu k^\mu +m_{3/2})\left[
-\eta_{\rho\sigma}+{1\over 3}\gamma_\rho\gamma_\sigma\right.
+\left.{1\over 3m_{3/2}}
(k_\rho\gamma_\sigma-k_\sigma\gamma_\rho)+{2\over 3m_{3/2}^2}
k_\rho k_\sigma\right]
\end{equation}
In the limit of vanishing axion mass, the other contributions
cancel except the term coming from $1/m_{3/2}^2$ term,
\begin{equation}
\sum |T|^2={m_{\tilde a}^6\over
6M^2m_{3/2}^2}
\left(1-{m_{3/2}^2\over m_{\tilde a}^2}\right)^2
\left(1+{m_{3/2}^2\over m_{\tilde a}^2}\right)
\end{equation}
where the initial spin average and final spin sum have been
taken into account.
The above expression looks dangerous for unitarity because vanishing
gravitino mass gives a divergent result for $|T|^2$. However, this
potential danger does not happen.   $m_{3/2}$ comes
together with $M$, giving a supersymmetry breaking scale
$M_S^2=Mm_{3/2}$.  Therefore, the above expression has a coefficient
$m_{\tilde a}^4/M_S^4$.  The supersymmetry breaking scale sets the
difference of masses between superpartners.  The axino mass can be
raised at most to $M_S$ compared to the axion mass.  In most axino
models, the axino mass involves an additional loop factor
suppression; thus $|T|^2/m^2_{\tilde a}$ can never be of order
unity and the unitarity problem does not arise with the above
expression.

The axino--gravitino cosmology is severely restricted by the
requirement of successful nucleosynthesis.  It will be shown below
that the axino mass is constrained
in a small band,
\begin{equation}
88\ {\rm keV}\left({m_{3/2}\over {\rm eV}}\right)^{2/5}
\left(\Omega_0h^2\right)^{1/10}
\ <\ m_{\tilde a}\ <\ 47\ {\rm MeV}
\end{equation}
where the lower bound comes from $\tau_{\tilde a}
<$ (the age of the universe)
and the upper bound comes from the requirement of successful
nucleosynthesis.  A more reliable lower bound comes from considering
the present energy density of axions and gravitinos from axino
decay, in which case the axino mass falls in the 1 MeV range.
Since we are interested in
the dark matter problem, we will concentrate on this interesting
possibility after discussing the nucleosynthesis constraint.  The
axino mass is predicted to be in 100 GeV or keV region in
supergravity models with supersymmetry breaking in the hidden sector
and minimal kinetic energy term \cite{ckn}.  In general, there exist
other loop contributions and the kinetic energy term can be of
nonminimal form.  Furthermore, we adopt the supersymmetry breaking
by supercolor type interactions in this paper
\cite{dine}.  Therefore,
the axino mass in the 1 MeV region is a theoretical possibility.

New particles whose interactions are suppressed by either
$F_a$ or $M_{Pl}$ are photino, saxino, and axino.  Because the
lifetimes of these particles can fall in the nucleosynthesis epoch,
we must consider the cosmological effects of these particles
carefully.

For $m_{\tilde\gamma}
>20$ GeV, the ratio of the photino number density to the
photon number density at the time of photino decoupling is less than
$10^{-9}$ as shown in Eq. (1), and we can neglect the effect of
photino.

Because we are interested in the axino--gravitino cosmology, we must
treat the supergravity interaction consistently. Then the saxino
cosmology weaken the axion energy crisis problem somewhat
\cite{kim}, due to
the photon reheating by the saxino decay.  For our eyeball
number $F_a=10^{12}$ GeV, the saxino mass determines the possibility
of closing the universe by cold axions.

Axinos come from two sources: thermal relic and decay products of
photino.  Because the photino number density is extremely small,
the axino is considered to be hot relic which decouples at
\begin{equation}
T_{\tilde aD}=10^{11}\ {\rm GeV}\ \left({F_a\over 10^{12}\
{\rm GeV}}\right)^2\left({0.1\over \alpha_c}\right)^3
\end{equation}
The ratio of the axino number density to the entropy density is
\begin{equation}
Y_{\tilde a}={45\zeta(3)\over 2\pi^4}{g_{\rm eff}\over g_{*s}(
T_{\tilde aD})}
\end{equation}
where $g_{\rm eff}\simeq 3/2$ for a two component axino.
Hence the axino number density is
\begin{equation}
n_{\tilde a}(T)=s(T)Y_{\tilde a}={\zeta(3)\over \pi^2}
{g_{\rm eff}g_{*s}(T)\over g_{*s}(T_{\tilde aD})}T^3
\end{equation}
As temperature goes down ($T<0.37m_{\tilde a}$), the axino
becomes nonrelativistic, and its energy density is given by
\begin{equation}
\rho_{\tilde a}(T)= m_{\tilde a}
n_{\tilde a}(T)={\zeta(3)\over \pi^2}
{g_{\rm eff}g_{*s}(T)\over g_{*s}(T_{\tilde aD})}m_{\tilde a}T^3.
\end{equation}
The axino dominates the energy density of the universe below $\tilde
T_{\tilde a}$ where
\begin{equation}
\tilde T_{\tilde a}\ =\ 2.4\ {\rm keV}\ \left( m_{\tilde a}\over {\rm MeV}
\right) \left( {930/4 \over g_{*s}(T_{\tilde aD})}\right)
\end{equation}
which corresponds to the cosmic time
\begin{equation}
\tilde t_{\tilde a}\ =\ 2.3\times 10^5\ {\rm sec}
\left({{\rm MeV}\over m_{\tilde a}}\right)^2\left(
{g_{*s}(T_{\tilde aD})\over 930/4}\right)^2\left(
{3.36\over g_{*}(\sim \tilde T_{\tilde a})}\right)^{1/2}
\end{equation}

A simple estimate of the upper bound of the axino mass is obtained
by requiring that the axino energy density at the time of neutrino
decoupling ($T_{\nu D}$) should be less than the energy
density of a single neutrino species:
\begin{equation}
m_{\tilde a}\ <\ {7\pi^4\over 120\zeta(3)}{g_{*s}(T_{\tilde aD})\over
g_{\rm eff}g_{*s}(T_{\nu D})}T_{\nu D}\simeq 47\ {\rm MeV}
\end{equation}

If the above axino mass bound is satisfied, the axino energy
density will
become the hot axion and hot gravitino energy densities now.  These
decay products are considered to be {\it hot} because
these are considered to be relativistic now.
On the other hand,
the coherent classical axion oscillation created at
$T=$ 1 GeV is responsible for the familiar
{\it cold} axion energy density \cite{pww}.

If cold axions do not close the universe, the hot axions and
gravitinos from axino decay can be made to close the universe
by choosing $m_{
\tilde a}$ and $m_{3/2}$ appropriately.
This is obtained from the present
ratio of the critical energy density to the radiation, $\rho_c/\rho_R
=h^2\ 2.3\times 10^4$ where $h$ is the Hubble parameter in units
of 100 km/sec/Mpc.  For these eyeball numbers of $m_{\tilde a}$ and
$m_{3/2}$, gravitinos are relativistic, and the age of universe
turns out to be too young.  By raising the gravitino mass to 10 keV,
it can be made that gravitinos become nonrelativistic and the universe
can be made matter dominated again quite lately and still the age problem
can be circumvented, e.g. even for the matter domination commencing from
$z=10$.  This scenario is a possibility, but the argument is intreaging,
and we do not consider this possibility any more.

An easier solution to the age problem is obtained by requiring that cold
axions \cite{pww} are responsible
for the matter dominated epoch.  In the remainder
of this paper, we briefly discuss this interesting scenario.

Cold axions might have dominated the mass density of the universe
sometime between $z=1,000$ and $z=10$ where the latter bound
is obtained to have a sufficiently old universe.  If cold axions
dominated the mass density of the universe at $z=10^3$ and axinos
decayed at $z=10^3$, the present gravitino energy density is
$\sim 10^{-3}$ times $\rho_{a}$ and the photon energy density
is $\sim 10^{-6}$ times $\rho_{a}$, which
is contrary to observation.  Therefore, cold axions might have
dominated the mass density of the universe quite lately.  We may take
$z=10$ for this purpose.  Then, at the time of axino
decay the cold axion energy density is
not important compared to the axino energy density.
Thus at the time of galaxy formation, the energy
densities of axinos (cold) and its decay products,
axions and gravitinos (hot), are most important.  If axino decayed much
earlier than the time of photon decoupling, the dark matters at the
time of galaxy formation were hot.  If axino decayed much later
than the time of photon decoupling at
$t_{\gamma D}$, the dark matters were cold axinos
at the time of galaxy formation.  Thus, by adjusting
the axino lifetime, one can introduce an appropriate ratio of hot
and cold dark matters at the time of galaxy formation.  For example,
$\tau_{\tilde a}\simeq 4.5 t_{\gamma D}$ gives 20\% and 80\% of
hot and cold dark matters, respectively, at the time of galaxy
formation.  In our case, $t_{\gamma D}=6.8\times 10^{11}
(\Omega_0h^2)^{-1/2}$ sec which
is  different from the standard estimate \cite{kolb} of $5.6\times
10^{12} (\Omega_0 h^2)^{-1/2}$ sec since in our case the universe
is radiation dominated from $t=\tau_{\tilde a}$ to $t=t(z=10)$.  A
0.72 MeV axino and a 1 eV gravitino satisfies the above lifetime
requirement,
\begin{equation}
\left({m_{\tilde a}\over 0.72\ {\rm MeV}}\right)^5\left(
{{\rm eV}\over m_{3/2}}\right)^2\
=\ 2h\Omega_0^{1/2}.
\end{equation}
Of course, a detailed simulation programmed for our new
suggestion is needed to obtain a realistic
ratio for the hot dark matter to the cold dark matter
at the time of galaxy formation in our scenario.

In conclusion, it is pointed out that axinos and light gravitinos
can influence the history of the universe quite nontrivially
if the axino and gravitino masses fall in the 1 MeV and
eV regions, respectively.  The axino lifetime of order of
$10^{12}$ seconds can provide both cold and hot dark matters
at the time of galaxy formation.

\centerline{\Large Acknowledgments}

This work is supported in part by KOSEF through
Center for Theoretical Physics,
Seoul National University, and by KOSEF--DFG Collaboration program.

\end{document}